*Original Article*

# Design and Production of an Autonomous Rotary Composter Powered by Photovoltaic Energy

FATIMA ZAHRA SITI [#1], MUSTAFA ELALAMI[*2], FATIMA ZAHRA BERAICH[3] ,MOHA AROUCH[*4] , SALAH DINE QANADLI[*5].

[1,2,3,4]*Engineering, Industrial Management & Innovation Laboratory IMII, Faculty of Science and Technics (FST), Hassan 1st  University of Settat, Morocco*

[1]f.siti@uhp.ac.ma,

[5]*Cardio Thoracic  and Vascular Unit, Department of Radiology, University Hospital of Lausanne, Bugnon 46, 1011 Lausanne, Switzerland,* Salah.Qanadli@chuv.ch,

*Abstract* -The problem of household waste management is becoming more and more acute with the growth of economic development that the country of Morocco has experienced. Moreover, the management of this waste is a burden for the municipalities, in view of its cost, which is increasing with time. Therefore, in this work we present the design, the mechanical and photovoltaic study of a new autonomous solar composter intended mainly for households. It allows to transform organic waste in situ, into a good quality compost that serves as a soil conditioner, in a short time compared to other composting systems, these times do not exceed 4 weeks. This innovative technology will reduce the amount of waste going to final landfill, or incineration, and exploit the compost produced in gardening and horticulture, which will be a very effective solution for waste management in Morocco.

## I. INTRODUCTION

In Morocco, more than 5.2 million tonnes of household waste are produced per year, almost all of which is disposed of either in landfills or incinerated [1]. However, the improper disposal of biodegradable waste represents a great risk to the environment, often difficult to control, such as: consumption of space, production of leachate, risk of groundwater contamination, not to mention the problems related to the emission of greenhouse gases [2]. Therefore, the use of composting, which is an ancestral method, presents an efficient and cheaper solution to reduce bio-waste at the source, before it is sent to final disposal [3], [4].  It offers people the opportunity to improve their economic conditions through backyard gardening, marketing of their compost and recyclables [4]. Composting is defined as a process of biological degradation of organic matter by microorganisms, including bacteria and fungi, which survive under specific conditions of temperature, oxygen, and humidity [5]. It significantly reduces the volume of organic waste because the process converts a large amount of biodegradable waste to carbon dioxide gas and water. Thus, the optimal conditions for composting are: aeration, moisture content between 40% and 65%, temperature between 43°C and 65°C, carbon to nitrogen ratio, and particle size [4], [6].

Known composting technologies are either open, closed, or reactor-based. In particular, composting in reactors, vermicomposter, or rotating drums remains the best option, as it eliminates unpleasant odours, accelerates the treatment process, and provides a favourable environment for the biodegradation of organic matter [7]-[9]. However, there is a multitude of composting methods, which is still unfortunately inefficient due to the penetration of harmful elements into the environment, and which affect the composting parameters [10]. The issue of controlling these parameters is becoming more important, especially with the use of rotary drums, composting is becoming more controlled and efficient, in terms of producing good quality compost in a short time [5], [11]-[13]. The rotary drum composter is a decentralised composting technique that provides agitation, aeration, and mixing of the substrate to produce mature and stable compost.  It reduces the composting time to 2-3 weeks. Different types of waste can be composted effectively in the rotating drum [14], [15]. In the relevant literature, various types of composters have been proposed. Among them, a closed rotary composter was developed in [16], it has a capacity of 50kg, and it allows to optimise the composting time in 2-3 weeks, this composter turns manually, with a frequency of 2 times a day. Another rotating composter was proposed in [15], it is equipped with an electromechanical system, which allows to facilitate the operation of this composter. In addition, an improvement of the operating system of the rotary composter, and a remote control system of the composting process based on the use of sensors and a man-machine interface was studied and presented in [17]. In the same work developed in [18], the authors designed and studied an autonomous rotary composter, equipped with a remote management system that allows the monitoring of the composting parameters from a distance, using a smartphone. However, during the operation of the composter, they discovered :

- A misalignment problem between the two rolling bearings,

- A blockage of the gear motor due to the high load applied to the worm gear,

- Wear on the cylinder drive shaft,

- Clavity matting.

To solve these problems, the authors in [19], have made improvements to the design, operation of the composter, and the control part that allows monitoring the composting parameters remotely, and in real time. However, this composter remain alimented by electrical energy. The exploitation of solar energy in powering systems has several advantages, such as cost reduction, energy saving, and environmental conservation. This solar energy had been used to power many systems such as in [20]-[22].

The work presented in this paper exploits the study carried out in [18] and [19], to design and implement an autonomous, domestic, and solar composting system. The objective of this study is to present a rotary composter, which represents an effective and innovative solution for the management of household waste, in fact this composter is autonomous, it is equipped with a system of control and remote monitoring, and in real time, so the main contribution in this work is illustrated at the level of the exploitation of photovoltaic energy, in the power supply of the operating system of the innovative composter, which characterizes this composter compared to the the works quoted above, and which shows the added value. The structure of this work, will be presented as follows: the first part will be intended for the description of the machine, and the design, then in the second part the SYSML analysis will be presented, in the third part the mechanical study will be proposed, and in the fourth part will present the photovoltaic study of the system, then the results and discussion of this work, and finally the conclusion.

## II. MATERIALS AND METHODS

### A. Composter Description

The composter is a machine in the form of a cylindrical drum, with a capacity of 0.181 m³, length 640 mm and diameter 600 mm, which allows household waste to be composted within four weeks, thanks to an aerobic fermentation process. The composter is constructed from a rigid, weatherproof plastic material (the exact nature of the material) and is supported by a frame, the legs of which are 680 mm long Fig.1.

The loading of the organic waste into the composter and the emptying of the compost is done through two doors located on the longitudinal side of the drum (length 225 mm and width 195 mm) Fig.1. The aeration of the composter is achieved through the two air vents located on both sides of the drum Fig.1, this aeration helps to avoid bad odours.

In addition, the composter contains a cylindrical shaft which allows it to be fixed to the stand, and is guided in rotation by a pulley-belt transmission system, and rotates at a speed of 4 rpm, thanks to a direct current electric motor and a parallel gear reduction unit. The drive system of the composter is powered by direct current from a photovoltaic system consisting of a solar panel, a regulator, and a solar battery.

The composting parameters in the rotating drum, namely temperature, humidity, NH3 gas and PH, are monitored by wireless sensors. The collected data is then communicated through an Arduino UNO board, which allows remote monitoring of the composting parameters.

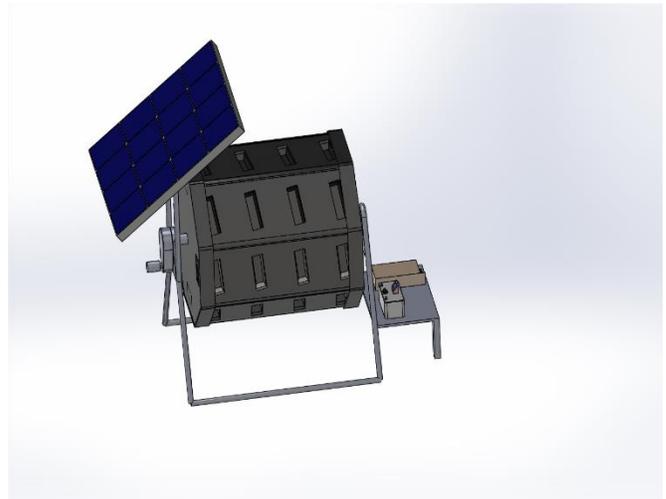

**Fig 1 Assembly diagram of the rotary composter**

### B. SYSML analysis

#### 1) composter requirements Diagram

The composter transforms organic waste into good quality compost in a short period of time (2-3 weeks), the specifications of the composter are described in the diagram in                           Figure                          2.

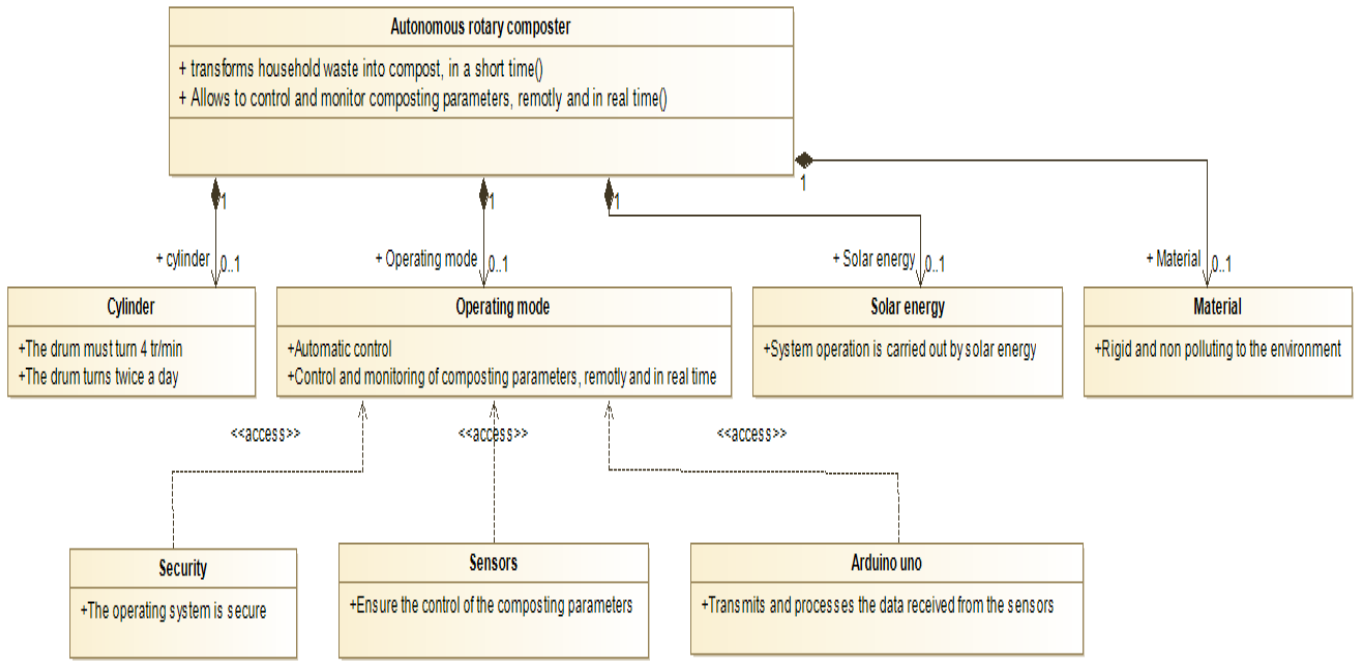

*Fig2 : Requirements diagram of the autonomous rotary composter*

**2) Block definition diagram of the composter**

The composting system consists of the following components:

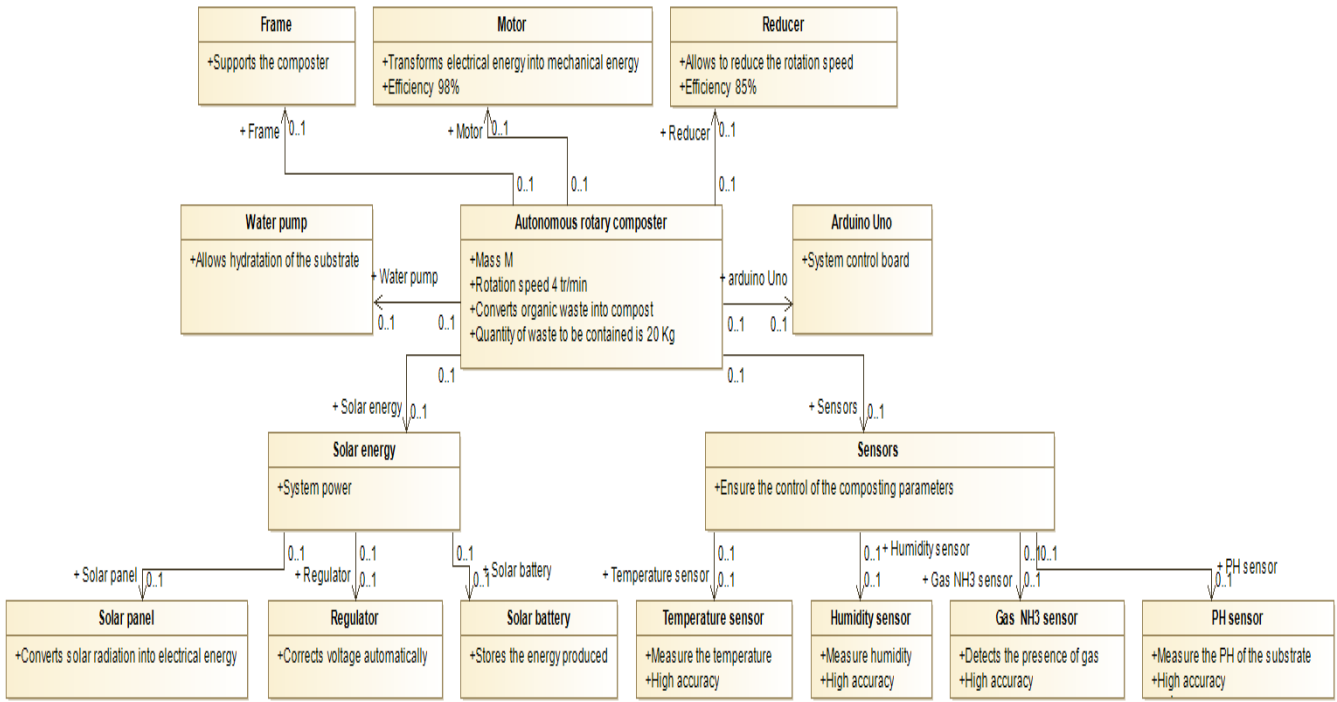

*Fig 3: Block definition diagram of the composter*

### 3) Internal block diagram

Used to describe the flows between blocks:

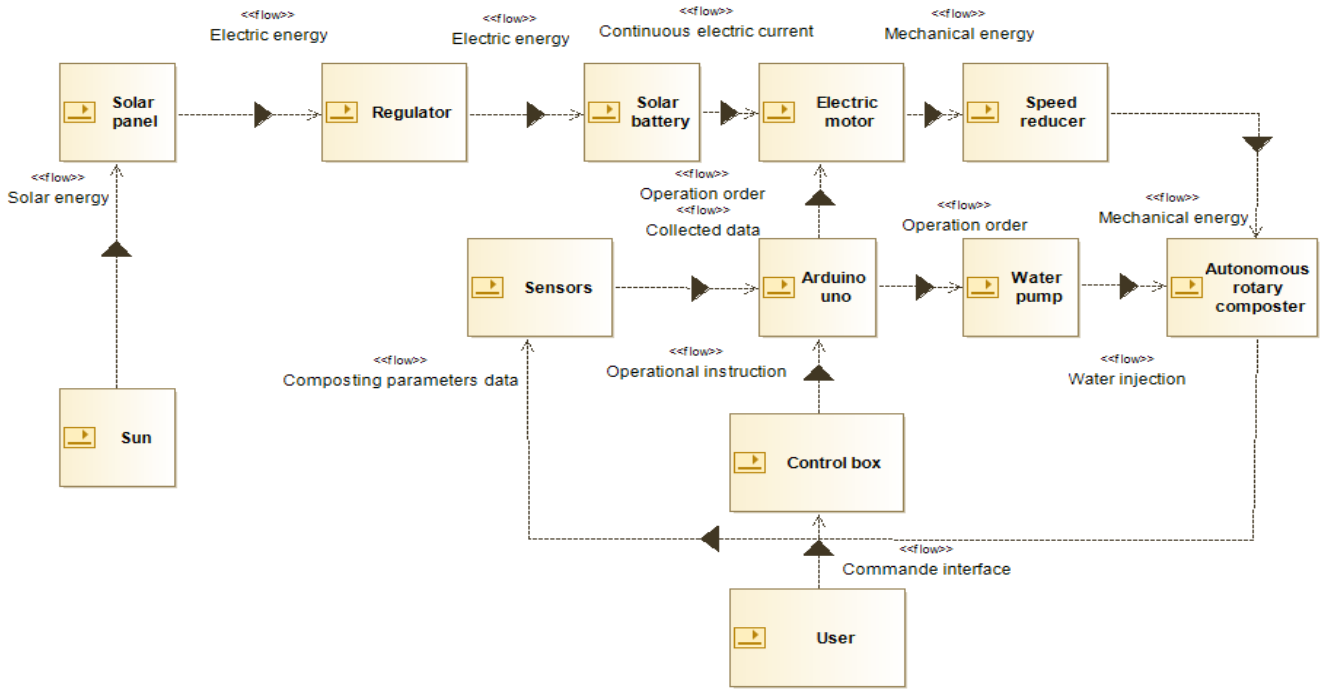

**Fig 4: Internal block diagram of the composte**

### C. Sizing the composter

#### 1) Composter Characteristics

The composter is in the form of a closed hollow cylinder, with dimensions (length L, diameter D, thickness e). It rotates at a speed of 4 rpm, has a mass M, and is made of PVC.

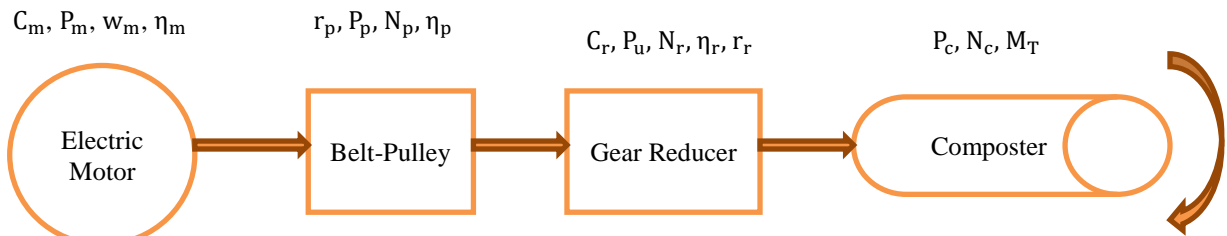

**Fig 5: composting system information flow**

#### 2) Determination of the Drive Torque

The rotary composter is in the form of a hollow cylinder of mass M, it can contain a quantity of waste of mass m, and it rotates with a speed of 4 rpm.

To make the composter rotate, the driving power of the composter must be equal to the reducer power, i.e.

$P_c = P_u$ (1)

Or $P_c = M_T \times g \times W_c \times r$ (2) et $P_u = C_r \times W_r$ (3)

Therefore $C_r = M_T \times g \times W_c \times r / W_r$ (4)

Thus, the rotational speed of the composter is slow (4rpm) compared to the motor speed, so the speed reducer needed to reduce this rotational speed must have a reduction ratio i determined by the following formula:

$i = W_m / W_r = \frac{P_m / C_m}{P_u / C_r}$ (5)

### 3) Determination of the Motor Torque

The motor torque required to drive the rotary composter is calculated by the following formula:

$C_m = \frac{C_r}{i \times \eta_m \times \eta_r}$ (6)

Knowing that: $\eta_m = \frac{P_p}{P_m}$ (7) and $\eta_r = \frac{P_u}{P_p}$ (8)

Such that:

r: radius of the composter

$P_c$ : Composter power

$M_T$ : Total mass (mass of composter + mass of waste)

g : gravitational constant

$W_m$ : rotation speed of the electric motor

$C_m$ : motor torque

$\eta_m$ : motor efficiency

$P_m$ : motor power

$C_r$ : gearbox torque

$W_r$ : speed of the gearbox

i : total reduction ratio

$\eta_r$ : gearbox efficiency

$P_p$ : power at the output of the pulley-belt

$P_u$ : output power

### 4) Transmission and reduction of motion

The composter rotates at a speed of 4 rpm, and since the speed of the electric motor is 1500 rpm, we decided to install a speed reducer (reduction ratio $r_r$, and efficiency $\eta_r$), and a belt-and-pulley transmission system (reduction ratio $r_p$, and efficiency $\eta_p$).

So the power output of the electric motor is given by :

$P_u = \frac{P_m}{\eta_r \times \eta_p}$ (9)

### Sizing of the belt pulley

The belt pulley is sized according to the standards used in [18]. The distance between the two axes is C = 110 mm. The diameter of the driving pulley is $d_m = 15$ mm. The diameter of the driven pulley is determined by the following formula: $d_r < C < 3(d_m + d_r)$ (10)

The ratio of reduction is : $r_p = \frac{d_m}{d_r}$ (11)

To determine the length of the belt, we must first specify the following angles:

$\theta_1 = \pi - 2\beta$ (12)

$\beta = \sin^{-1}(\frac{d_r - d_m}{2C})$ (13)

$\theta_2 = \pi + 2\beta$ (14)

$L = \sqrt{4C^2 - (d_r - d_m)^2} + \frac{1}{2}(d_m\theta_1 + d_r\theta_2)$ (15)

### Gearbox sizing

The parallel gear reducer reduces the speed of the electric motor and increases the force. Its reduction ratio is determined by the following formula: $r_r = \frac{N_c}{r_p \times N_m}$ (16)

### 5) Photovoltaic study of the composter

To determine the electrical energy needs for the composter, it is necessary to determine the power of the photovoltaic generator, the number of panels, the regulator, and the associated battery capacity to be installed.

### Energy consumed:

The composter has to turn twice a day, and each turn lasts 10 min, so the number of hours of operation of the composter per day is about 20 min/day.

Therefore, the energy that will be consumed per day by the composter is such as:

$E_c = P_m \times$ number of hours of operation composter (1)

### Energy produced:

To ensure that the energy requirement of the composter is met, the energy produced by the photovoltaic system must satisfy the following relationship:

$E_p = \frac{E_c}{K}$ (2)

Moreover, since the photovoltaic system contains a battery, the K-factor is generally equal to 0.65.

### The solar panel:

Determining the number of solar panels to be used is essential for photovoltaic sizing, therefore we must first determine the peak power of the solar panel, then:

$$P_c = \frac{E_p}{I_r} \quad (3)$$

knowing that : $I_r$=5 Kwh/$m^2$/jour

*The solar battery :*

To store the electricity produced by the photovoltaic panel, in order to use it at the right time, it is necessary to choose the appropriate choice.

So to store the daily energy need of the composter (221 < 1000 wh), we can choose a 12 V battery, with a 3 days autonomy.

As for the storage capacity of the battery, it will be calculated by the following formula:

$$C = \frac{E_c \times N}{D \times U} \quad (4)$$

Thus the amperage of this battery is such that

$$I_{battery} = U \times D \quad (5)$$

In addition, the required number of batteries is such that:

$$n_{battery} = \frac{I_{battery}}{C} \quad (6)$$

*The regulator:*

The choice of controller for this photovoltaic installation must satisfy the following conditions:

$$U_{regulator} = U_{generator} \quad (7)$$

$$P_{regulator} > P_{crete} \quad (8)$$

$$I_{regulator} > \frac{P_{crete}}{U_{generator}} \quad (9)$$

Knowing that:

N: battery life

C: storage capacity of the battery

D: the discharge of the battery (equal to 0.9 since the battery is made of lithium)

U : battery voltage (12 V)

$E_c$: Daily energy consumed by the composter

$I_{battery}$: The battery amperage

$E_p$: Energy produced

$U_{regulator}$: The voltage of the regulator

$I_{regulator}$ : The amperage of the regulator

$P_{regulator}$: The power of the regulator

$U_{generator}$ : The voltage of the photovoltaic generator

$P_{crete}$ : The peak power of the photovoltaic panel

### III. RESULTS AND DISCUSSION

#### A. Case study: Autonomous solar rotary composter and component selection

According to the above mechanical and photovoltaic study, to drive the solar rotary composter, we need components whose characteristics are as follows:

**TABLE I :** *Components characteristics*

| Component | characteristics |
|---|---|
| Motors | DC electric motor : <br> Cm = 2,74 N.M <br> Nm = 1500 tr/min <br> $\eta_m$= 98% |
| Reducer | Gear reducer: <br> Cr = 89,7 N.M <br> Nr = 4 tr/min <br> I = 39,27 <br> $\eta_r$= 85% |
| Organic waste | The amount of organic waste that can fit in the composter is 20 Kg |
| Photovoltaic panel | Polycrystalline photovoltaic panel : <br> Nbr = 1 <br> $P_{crete}$ = 20 watt (Maximum Power (W)) <br> Voltage at maximum power point Vmp (V) = 17,1 V <br> Current at maximum power  Imp(A) = 1,17 A <br> Open circuit voltage  Voc (V) = 21,3 V <br> Short-circuit current  Isc (A) = 1,31 A |
| Regulator | The voltage regulator : <br> $U_{regulator} = 12\ V$ <br> $P_{régulateur regulator} > P_{crete} = 20\ watt$ <br> $I_{regulator} > \frac{P_{crete}}{U_{generator}} = 1,67$ Amp |
| Solar battery | $U_{Battery}$= 12 V <br> N = 3 days <br> $I_{battery} = 6,6\ Amp$ <br> C = 40 Ah <br> $n_{battery} = 1$ |

The realization of this work is the fruit of a successful collaboration with the company Biodôme, and with the team of the laboratory of engineering, industrial management and innovation. During this study, a rotating, autonomous, and solar-powered composter was designed and constructed. It allows to compost household waste, to monitor and control the composting parameters remotely and in real time, thanks to a remote management system, so the operation of this composter is based on the exploitation of solar energy.

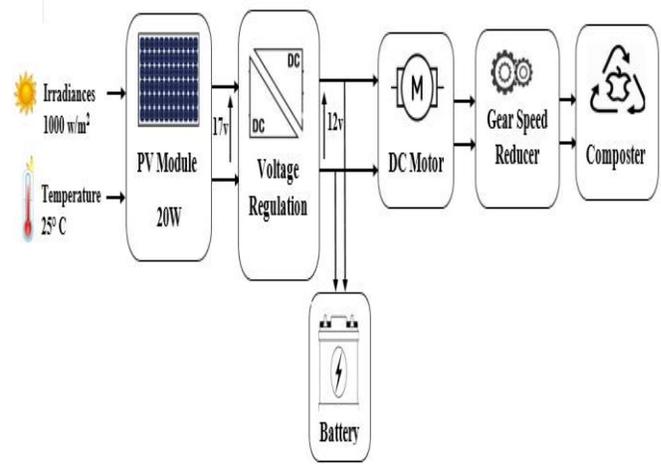

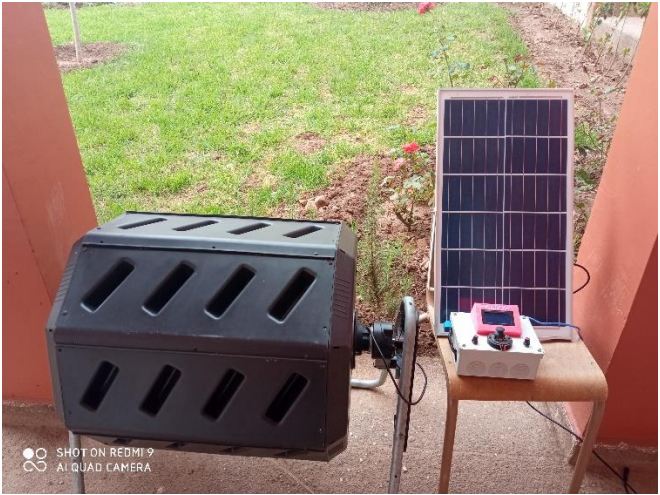

*Fig 6: Reel solar composter*

## B. Simulation of results

### 1) Photovoltaic system

Solar energy is one of the most used renewable energies, so in order to optimise the electrical energy needed to power this system we decided to exploit the electrical energy converted from solar radiation, thanks to the photovoltaic system.

*Fig 7: Block diagram*

The figure 7 present the photovoltaic system block diagram. A PV module witch represents the energy source with a maximum power of 20W. This panel generates a maximum voltage of (17.1V) and a maximum current of (1.17 A). The second block represents a voltage regulation DC-DC that gives in its output a 12V. The main role of the static converter is to ensure impedance matching so that the photovoltaic panel PV delivers maximum energy. The DC Motor used is alimented with a 12v voltage. A gear speed reducer is used to reduce the speed of our motor to rotate the Composter studied.

The diagram shown in Figure 7 is represented in MATLAB Simulink in Figure 8.

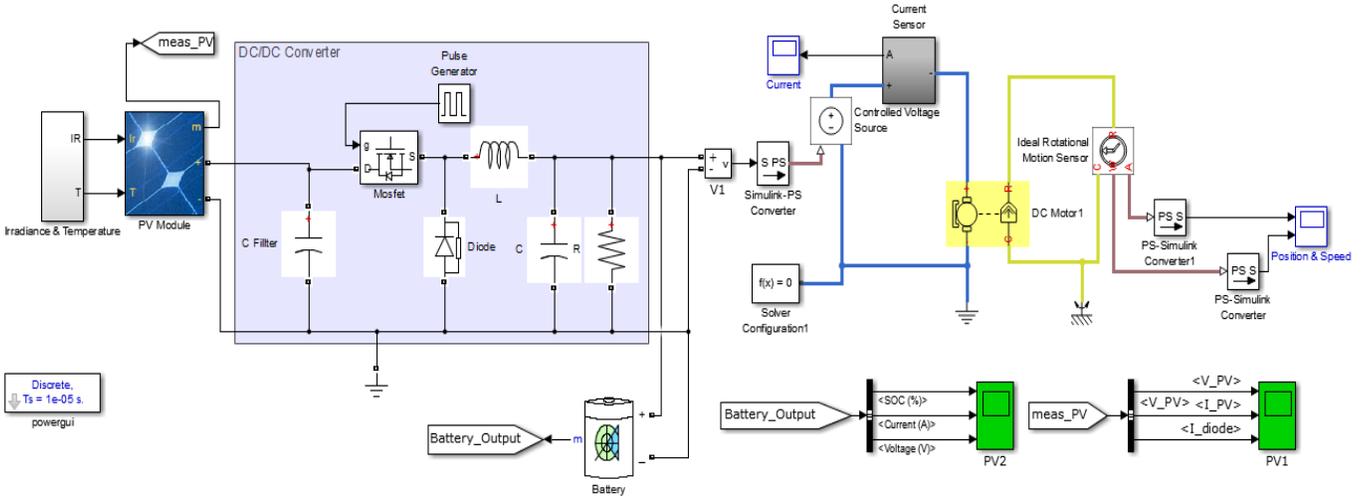

**Fig 8: Simulink model of the proposed system**

The results of this simulation are shown in Figures (9-12).

### 2) characteristics of the PV Module

The Figure 9 display the I-V and P-V characteristics of the PV Module used for variable temperatures (25°C and 45°C) with an irradiance of 1000W/m². It's can be clearly seen from this figure that the maximum power point highly dependent on the temperature value.

The PV solar module used in this study consists of 60 polycrystalline silicon solar cells electrically configured as five series strings. Its main electrical specifications are shown in Table I.

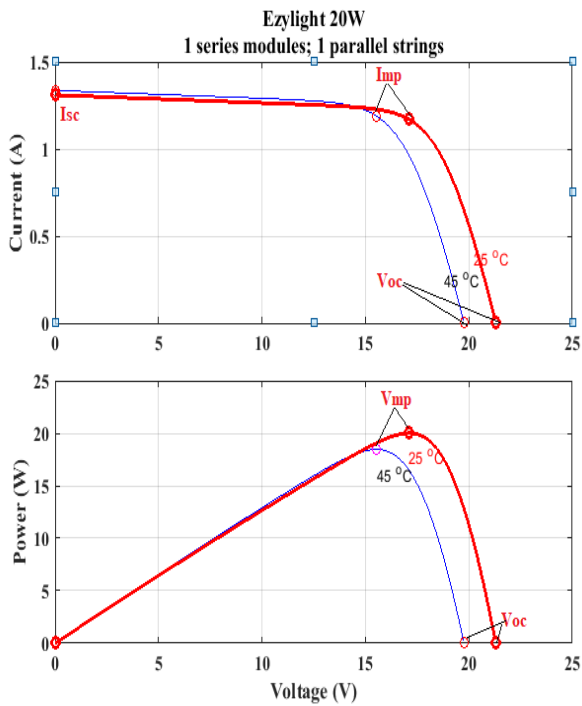

Fig 9: PV Module characteristic I=f(V) for two different Temperature

The simulation results shown below are taken at a temperature of 25°C, and radiation varying between 800 W/m² and 1000 W/m². The PV module characteristic are mentioned in TABLE II.

**TABLE II :** *PV characteristics for a fixed irradiance (1000w/m²) and varied temperature*

| T(°c) | Im(A) | Vm(V) | Pm(W) |
|-------|-------|-------|-------|
| 25 | 1.170 | 17.10 | 20.01 |
| 45 | 1.174 | 15.55 | 18.26 |

The figure 10 presented below illustrate the characteristics of the PV solar module for differents values of radiation and a fixed temperature at 25°C. I_PV, V_PV and P_PV represent respectively the currant, the voltage and the power obtained. When the radiation rate increases, Isc increases and Vco decreases slightly, therefore, the maximum power undergoes a significant increase.

The maximum power obtained is about 20w correspond to an irradiance of 1000W/m² as described in the PV Module datasheet. The maximum currant is about 1.17A and the maximum voltage is equal to 17.1V. TABLE III.

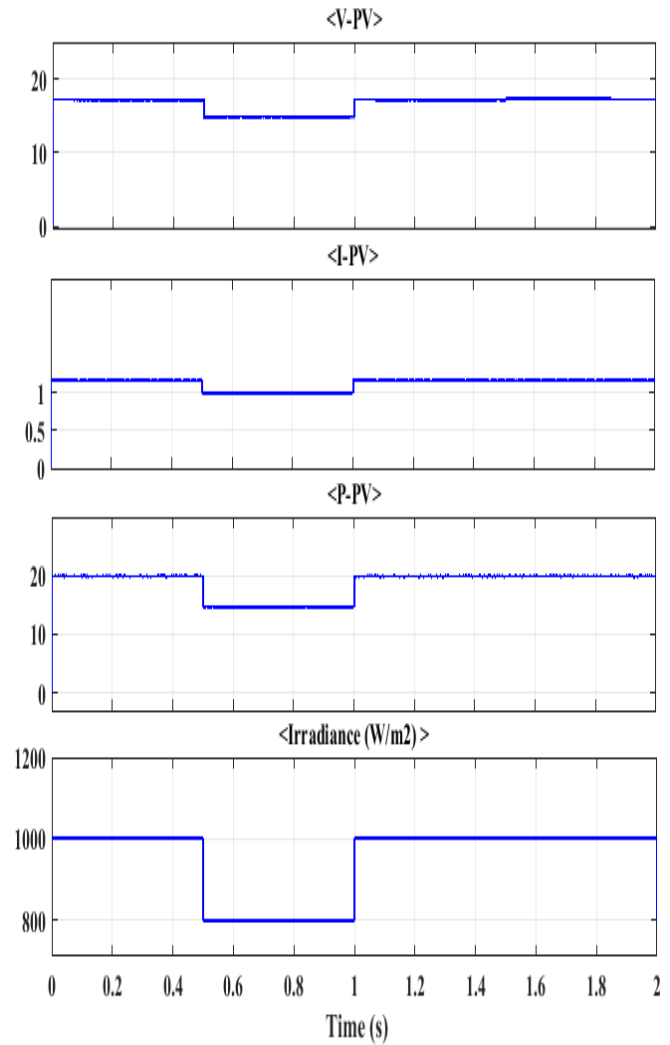

Fig 10: PV module output with varying sun irradiance in W/m2

**TABLE III :** *PV module characteristics for a fixed temperature (25•c) and varied sun irradiance*

| Irr(w/m²) | Vpv(V) | Ipv(A) | Ppv(W) |
|-----------|--------|--------|--------|
| 1000 | 17.30 | 1.16 | 20.03 |
| 800 | 14.78 | 0.99 | 14.73 |

The figure 11 illustrate the battery state of charge when a voltage of 12v is applied. Initially the battery is on 45% of charge.

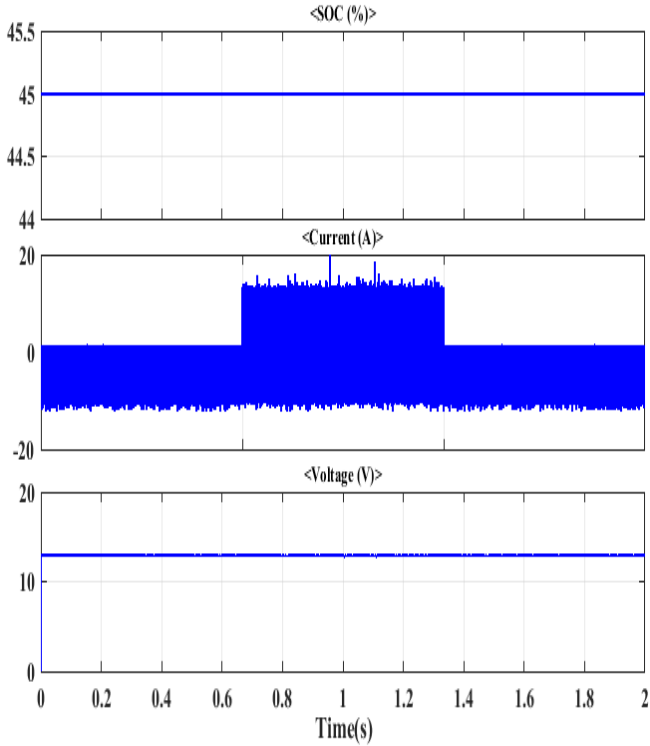

*Fig 11: Battery state*

The figure 12 represent the DC Motor position and speed when a voltage of 12v is applied. From the 0.5s, the DC Motor speed stabilizes at 150 rpm.

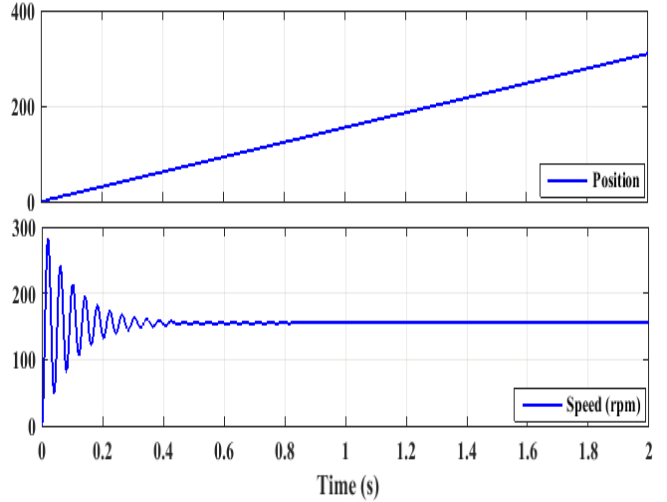

*Fig 12: Motor speed and position*

The results of the simulation in MATLAB Simulink, proved that the photovoltaic system used gives sufficient and efficient energy to run the composter.

## VI. CONCLUSIONS

This paper contains the design, the mechanical study, and the photovoltaic study of a rotating, autonomous, and solar composter. Indeed, this composter allows the transformation of household waste, which contains a large organic fraction, into a mature and stable compost, valid for gardening, horticulture, etc., and this within a period of no more than 4 weeks. In addition, the proposed composter present an innovative solution, which allows to improve the composting process, thanks to a remote management system that offers the control and monitoring of the composting parameters at a distance and in real time. Thus, the operation of the composting machine is based on the exploitation of solar energy, through a photovoltaic panel. the designed system has been successfully tested and simulated in Matlab Simulink enviroment. The simulation results demonstrate the design and the parameters of the proposed system. The added value is essentially manifested through the use of photovoltaic energy to power the composting machine. As a perspective of the present work, a maximum energy exploitation of the PV will be performed, while using MPPT control to adjust the duty cycle of the DC/DC converter. In addition, a work will be carried out on the choice of the manufacturing material, so that this material must have the following characteristics: environmentally friendly, commercially available and offering a good quality/price ratio.

## REFERENCES

[1] https://www.medias24.com/2018/11/05/dechets-menagers-au-maroc-moins-de-10-sont-revalorises/.

[2] Divyabharathi, R. Design and Development of In-Vessel Composter for Treating Agricultural Weeds.

[3]    Liu, Z., Wang, X., Wang, F., Bai, Z., Chadwick, D., Misselbrook, T., & Ma, L. The progress of composting technologies from static heap to intelligent reactor: Benefits and limitations. Journal of Cleaner Production, 270, 122328, (2020).

[4]    Jr, M. N. L., & Boado, M. M. M. Performance Evaluation of a Non-Odorous Compost Barrel for Household Purposes.

[5]    Hemidat, S., Jaar, M., Nassour, A., & Nelles, M. Monitoring of composting process parameters: A case study in Jordan. Waste and Biomass Valorization, 9(12), 2257-2274. (2018).

[6]    Guria, N., & Datta, S. DOMESTIC WASTE MANAGEMENT: HOME COMPOSTING BINS AND UTILIZING FOOD PROCESSING WASTE.

[7]    Alkarimiah, R., & Suja, F. Composting of EFB and POME Using a Step-Feeding Strategy in a Rotary Drum Reactor: The Effect of Active Aeration and Mixing Ratio on Composting Performance. Polish Journal of Environmental Studies, 29(4). (2020).

[8]    Makan, A., & Fadili, A. Sustainability assessment of large-scale composting technologies using PROMETHEE method. Journal of Cleaner Production, 261, 121244. (2020).

[9]    Arumugam, K., Seenivasagan, R., Kasimani, R., Sharma, N., & Babalola, O. Enhancing the post consumer waste management through vermicomposting along with bioinoculumn. Int. J. Eng. Trends Technol, 44, (2017). 179-182.

[10]   Gopikumar, S., Tharanyalakshmi, R., Kannah, R. Y., Selvam, A., & Banu, J. R. Aerobic biodegradation of food wastes. In Food Waste to Valuable Resources (pp. 235-250). Academic Press. (2020).

[11]   Ahmadi, T., Casas, C. A., Escobar, N., & García, Y. E. Municipal organic solid waste composting: development of a tele-monitoring and automation control system (2020).

[12]   Shalavina, E., Briukhanov, A., Vasilev, E., Uvarov, R., & Valge, A. Variation in the mass and moisture content of solid organic waste originating from a pig complex during its fermentation. Agronomy Research, 18(S2), 1479-1486. (2020).

[13]   Zhou, X., Yang, J., Xu, S., Wang, J., Zhou, Q., Li, Y., & Tong, X. Rapid in-situ composting of household food waste. Process Safety and Environmental Protection, 141, 259-266, (2020).

[14]   Kalamdhad, A. S., & Kazmi, A. A. Effects of turning frequency on compost stability and some chemical characteristics in a rotary drum composter. Chemosphere, 74(10), 1327-1334, (2009).

[15]   Kalamdhad, A. S., Singh, Y. K., Ali, M., Khwairakpam, M., & Kazmi, A. A. Rotary drum composting of vegetable waste and tree leaves. Bioresource Technology, 100(24), 6442-6450. (2009).

[16]   Divyabharathi, R. Design and Development of In-Vessel Composter for Treating Agricultural Weeds

[17]   Ahmadi, T., Casas Díaz, C. A., García Vera, Y. E., & Escobar Escobar, N. A prototype reactor to compost agricultural wastes of Fusagasuga Municipality. Colombia. (2020).

[18]   Elalami, M., Baskoun, Y., Beraich, F. Z., Arouch, M., Taouzari, M., & Qanadli, S. D. Design and Test of the Smart Composter Controlled by Sensors. In 2019 7th International Renewable and Sustainable Energy Conference (IRSEC) (pp. 1-6). IEEE. (2019, November).

[19]   Elalami, M., Lahmadi, M. M., Siti, F. Z., Baskoun, Y., Arouch, M., & Beraich, F. Z. Innovative Design and Realization of a Smart Rotary Composter with a Remote Management System. International Journal, 8(7). (2020).

[20]   Alfonso, R. N., Leyesa, M. C., Lapiguera, D. M., Florencondia, N., & Subia, G. S. 'proposed design for framework management of cryptocurrency: Study of the World's first digital currency,''. Int. J. Eng. Trends Technol., 68(1), (2020). 57-63.

[21]   Kebede, A. Study, Design And Modeling Of A PV-Powered DC-Home With Optimizedstorage Appliance For Rural Electrification: The Case Of Tullo Gudo Island, Ethiopia (Doctoral dissertation, ASTU). (2020).

[22]   Prasad, N. L., Sri, P. U., & Vizayakumar, K. Life Cycle Assessment of a 100 kWp Solar PV-Based Electric Power Generation System in India. In Recent Trends in Mechanical Engineering. Springer, Singapore. (2020). (pp. 81-94).